\documentclass[a4paper,pre,preprint,showpacs,amsmath,amssymb]{revtex4}
\usepackage{graphicx}% Include figure files
\usepackage{dcolumn}% Align table columns on decimal point
\usepackage{bm}% bold math
\newcommand{\be}{\begin{equation}}
\newcommand{\ee}{\end{equation}}

\begin{document}

\title{Exceptional discretisations of the sine-Gordon equation}
\author{I.V. Barashenkov}
 \email{Igor.Barashenkov@uct.ac.za,igor@odette.mth.uct.ac.za}
\affiliation{Department of  Mathematics,
University of Cape
Town, Rondebosch 7701, South Africa}
\author{T.C. van Heerden}
 \email{tvanheerden@gmail.com}
\affiliation{Department of  Mathematics,
University of Cape
Town, Rondebosch 7701, South Africa}

\date{\today}

\begin{abstract}
Recently, the method of one-dimensional maps
was introduced as a means of generating exceptional discretisations of the
$\phi^4$-theories, i.e., discrete $\phi^4$-models which support
kinks  centred at a continuous range of positions relative
to the lattice. In this paper, we employ this method to 
obtain exceptional discretisations of the sine-Gordon equation
(i.e.  exceptional Frenkel-Kontorova chains). We also
use one-dimensional maps to construct a
discrete sine-Gordon equation supporting kinks moving with arbitrary 
velocities without emitting radiation.

\end{abstract}

\pacs{05.45.Yv}
%\showpacs

\maketitle
%\tableofcontents{}

\section{Introduction}
\label{Intro}

Discrete analogs of nonlinear evolution equations have been the
subject of intense investigation over the last 15 years. A great
deal of insight has been gained into the properties of 
the discretised nonlinear Schr\"odinger, Landau-Lifschitz,
Korteweg-de Vries, $\phi^4$- and other equations which were
originally introduced in the context of continuous nonlinear media.
As for the discrete sine-Gordon equation that we study in this paper, 
it preceded the appearance
of its continuum counterpart in the physics literature.
The equation dates back to 1938 when it 
was proposed by Yakov Frenkel and
Tatyana Kontorova to model stationary and moving crystal
dislocations \cite{FK}.

The original Frenkel-Kontorova model consisted of a chain of
harmonically coupled atoms in a spatially periodic potential:
\begin{equation}
{\ddot \theta}_n = \frac{1}{h^2} (\theta_{n+1} - 2 \theta_n + \theta_{n-1})
- \sin \theta_n.
\label{FK}
\end{equation}
Here $\theta_n$ is the position of the $n$-th atom in the chain,
 $1/h^2$ is a coupling constant, and the overdots indicate
differentiation with respect to time: ${\ddot \theta}= d^2
\theta/dt^2$. An alternative interpretation of Eq.\eqref{FK}
is that of a chain of torsionally coupled pendula, with 
$\theta_n$ being the angle the $n$-th pendulum makes with the vertical.  
Finally,  Eq.\eqref{FK} can be seen simply as a
discretisation of the sine-Gordon equation
\begin{equation} 
{\ddot \theta}= \theta_{xx} - \sin \theta,
\label{sG}
\end{equation}
which was conceived for the numerical simulation of
this partial differential equation.

Since  its original inception as the Frenkel-Kontorova
model, the discrete sine-Gordon equation \eqref{FK} 
 has reappeared in a great number of physical contexts,
including domain walls in ferro- and antiferromagnetic crystals,
charge-density waves in solids, crowdions in metals, 
vortices in arrays of Josephson junctions,
incommensurate structures in metals and insulators, and nonlinear excitations in
hydrogen-bonded molecules.  (See \cite{FK_reviews,Grif} for review and references.)
The equation has also been generalised in a variety of ways.
In the present paper we study, systematically,  two classes of such generalisations.
In the models of the first class the main part of the intersite coupling is still
harmonic, as in the original Frenkel-Kontorova model, but
in addition there is an anharmonic part of interaction arising
from the modified periodic potential. These types of models 
are of interest primarily in the stationary case where they define 
nontrivial systems of  
statistical mechanics (systems with convex interactions)
\cite{Aubry,Grif}.
The stationary discretisations in this class have
the form
\begin{equation}
 \frac{1}{h^2} (\theta_{n+1} - 2 \theta_n + \theta_{n-1})
= f(\theta_{n-1}, \theta_n, \theta_{n+1}),
\label{dsG1}
\end{equation}
where the function
$f$ (not necessarily a periodic function)
reduces to $\sin \theta$ in the continuum limit:
\[
f(\theta_{n-1}, \theta_n, \theta_{n+1}) \to  \sin \theta_n
\quad \mbox{as} \quad \theta_{n-1},  \theta_{n+1} \to \theta_n.
\]

The other class consists of all-periodic discretisations:
\begin{equation}
  \frac{1}{h^2} \sin (\theta_{n+1} - 2 \theta_n + \theta_{n-1})
= f(\theta_{n-1}, \theta_n, \theta_{n+1}).
\label{dsG2}
\end{equation}
Here $f$ is a periodic function of each of its three arguments, 
such that $f(\theta_{n}, \theta_n, \theta_{n})=\sin \theta_n$. 
Equations of this type govern arrays of electric dipoles
or magnetic spins in which the interactions between the neighbouring elements 
 are characterised by the trigonometric functions of the corresponding angles.
These models  are commonly referred to as the ``sine lattices" \cite{sine_lattices}. 
In the stationary case, examples of the sine lattices include the usual and the chiral
one-dimensional  $XY$ model in the magnetic 
field \cite{chiral_XY,Grif}. 
In the time-dependent setting, the sine lattices
were used to model the rotational dynamics of methyl groups in 4-methyl-pyridine
\cite{Fillaux}, $CH_2$ units in crystalline polyethylene \cite{TH1}
and bases in a DNA macromolecule \cite{TH1,TH2}; to study
conformational defects in polymer crystals \cite{ZCK}
and nonlinear waves in chains of electric dipoles \cite{Speight_Zolo}.

In addition to \eqref{dsG1} and \eqref{dsG2}, we also consider
some other
discretisations which have the special property of exhibiting exact solutions.

In most physical applications of the sine-Gordon theory, both continuum and discrete,
the central role is played by its solitary-wave solution, called a
 kink. The kink represents a dislocation in the crystal, a $2 \pi$-twist wave 
in the chain of pendula, and a quantum of magnetic flux  in a long Josephson junction.
In the continuum model \eqref{sG}, the kink solution is available explicitly;
in particular, the stationary kink has the expression
\begin{equation}
\theta(x)= 4 \, {\rm arctan} \, [ \exp \, (x-x^{(0)})].
\label{kink}
\end{equation}
The stationary kink \eqref{kink} depends on a single parameter, the position of its center $x^{(0)}$,
which can be varied continuously: $-\infty < x^{(0)} < \infty$.
In generic discretisations of the sine-Gordon theory, however, 
the kink can only be centred 
 on a lattice site or strictly midway between two neighbouring sites
\cite{Currie,Peyrard_Remoissenet,Ishimori_and_Peyrard,Boesch_Munakata_Flach}.
Mathematically, this is a consequence of the breaking of the 
translation symmetry of the continuum model. 
The physical interpretation is that 
the discrete kink 
can only remain stationary when placed at a minimum or a maximum of the so-called
Peierls-Nabarro barrier, a periodic potential induced by the discretisation 
of equation \eqref{sG} 
\cite{Currie,Peyrard_Remoissenet,Ishimori_and_Peyrard,Boesch_Munakata_Flach,Willis}.

Speight and Ward \cite{SW} were  the first to realise  that the breaking
of the translation symmetry does not necessarily preclude the existence of
a one-parameter family of stationary discrete kinks with an arbitrary centring 
relative to the lattice. In other words, despite not being translation invariant, 
the lattice equation may support a kink solution which depends on a 
continuous translation parameter. This ``spontaneous symmetry restoration"
is a nongeneric phenomenon which may only occur in isolated, or {\it exceptional},
discretisations of the sine-Gordon model. Physically, it implies that the discretisation does not
induce the Peierls-Nabarro barrier, or that the barrier is transparent to kinks.
Flach, Zolotaryuk and Kladko have discovered a similar phenomenon
in a class of discrete Klein-Gordon systems with  nonlinearities of a
special form \cite{FZK}.

The classification of the exceptional discrete sine-Gordon equations, i.e., equations
supporting families of stationary kinks with a continuously variable position 
relative to the lattice,
is of fundamental interest. Firstly, 
the discrete kinks tend to be more mobile in  
exceptional discrete models. There are some isolated velocities
at which kinks in the exceptional models may {\it slide}, i.e. travel without losing energy to radiation
\cite{OPB}.
 For other velocities,
the moving kinks do radiate but the amplitude of radiation is much smaller than 
in generic  systems \cite{SW,K0,K2,OPB}.
In addition, the collisions of kinks were reported to be more elastic 
in the exceptional models \cite{K3}.
We also show in this 
paper that some exceptional
systems admit time-dependent versions which 
support sliding kinks with {\it arbitrary\/} velocities.
Furthermore, there are indications \cite{DKKS}
that all exceptional discretisations possess a conservation law: they conserve either
 energy or momentum. Therefore, exceptional discrete models appear to be ``better" approximations 
of the partial-differential equation \eqref{sG} --- at least  as far as the kink solutions
are concerned --- as they preserve important properties of the continuum model.

The objective of the present paper is to identify exceptional Frenkel-Kontorova 
models within the families \eqref{dsG1} and \eqref{dsG2}. 
We are not going to attempt  a complete classification
here;
instead, we focus on identifying simple particular cases
which may be of practical use in future.
We also construct two discrete models 
with {\it exact\/} (stationary and moving) kink solutions.

An outline of the rest of the paper is as follows. In the next section
(section \ref{Method}), we present the method
of one-dimensional maps as applied to discrete sine-Gordon equations.
In section \ref{Mixed}, the method is used to identify simple 
exceptional discretisations of the form \eqref{dsG1} involving ratios of trigonometric and linear
functions. The symmetric maps found in this section have one
further use;  in section \ref{Pure} we utilise them to construct
purely trigonometric discretisations [of the form \eqref{dsG2}].
In the subsequent sections we present discrete sine-Gordon equations 
with {\it exact\/} stationary (section \ref{Exact}) and moving (section \ref{Travelling})
kink solutions. Finally, several concluding remarks are made in section \ref{Conclusions} which summarises the results
of this study.

\section{The method of one-dimensional maps}
\label{Method}

To derive an exceptional discrete sine-Gordon
model,
Speight and Ward used the Bogomolny energy-minimality argument \cite{SW}.
The 
energy minimality requirement has naturally led them
to consider a one-dimensional map  rather than the original, second-order,
difference equation. In the follow-up work \cite{Speight_1997}, Speight utilised the energy-minimising map
(the Bogomolny map)
 to prove the
existence of a one-parameter family of kinks
for their discretisation of the $\phi^4$-theory,
i.e. the exceptionality of their $\phi^4$-model.

A further insight was due to Kevrekidis \cite{K}.
Inspired by Herbst and Ablowitz' results on the
discrete nonlinear Schr\"odinger equation
 \cite{Herbst_Ablowitz}, Kevrekidis reformulated the
exceptionality of
a stationary discrete Klein-Gordon equation
as the existence of a two-point invariant. He also
 provided two
phenomenological recipes of construction of stationary
discretisations with such invariants.
Thus, the existence of
a two-point invariant replaced  the 
energy minimality requirement as the crucial property of 
exceptional discretisations.

The  universality of one-dimensional maps as generators of 
translationally-invariant families of solutions
has been fully realised
in Ref.\cite{BOP}. 
 Instead of trying to identify discretisations
 exhibiting a two-point invariant, the authors of \cite{BOP}
 proposed to generate exceptional discretisations
 departing from a postulated map. (That is,
 the two-point invariant has now become a starting
 point rather than the final objective of the analysis.)
 In this way, the classification of exceptional
 discretisations has been reduced to the classification
 of one-dimensional maps.  This will be our approach
in this paper as well.

We start by considering a discrete sine-Gordon equation of the form \eqref{dsG1}
and assume that the 
corresponding stationary equation,
\begin{equation}
\frac{1}{h^2} (\theta_{n+1} - 2 \theta_n + \theta_{n-1})
= f(\theta_{n-1}, \theta_n, \theta_{n+1}),
\label{stat}
\end{equation}
has a  solution of the form
$\theta_n=g(nh)$, where the continuous function
$g(x)$ is defined 
for $-\infty< x< \infty$ and is monotonically
growing, with $g(-\infty)=0$
and $g(\infty)= 2 \pi$. 
Since $n$ does not appear in equation \eqref{stat} explicitly, 
from the existence of the above solution
it follows that equation \eqref{stat} also 
has a whole family of solutions $\theta_n=g(nh-x^{(0)})$, with
any real $x^{(0)}$, and therefore, that the model \eqref{stat} is exceptional.
For each $x^{(0)}$, the solution $\theta_n=g(nh-x^{(0)})$
represents a discrete kink; if we interpret values $x_n=nh$ as positions of the 
lattice sites on the $x$-axis, the kink $\theta_n$ appears centred on the point $x=x^{(0)}$. 
It is important to emphasise that we do not need to know an explicit form
of $g$; all we need to know is that a function with these properties exists
(for example, as an implicit function).

As $g(x)$ is a monotonically growing function, we can invert it to
obtain $\theta_{n+1}=g(g^{-1}(\theta_n) +h) \equiv F(\theta_n)$. Since $g(x)$ is defined for all
real $x$, the function $F(\theta_n)$ is defined for any $\theta_n$.
 Thus the fact that the discretisation
is exceptional implies that 
the kink solution  $\theta_n$ satisfies a one-dimensional map \cite{BOP}.
The opposite is also true. Namely, assume  
Eq.\eqref{stat} results from the iteration of a one-dimensional map $\theta_{n+1}=F(\theta_n)$
 (in a similar
way as a second-order differential equation can be derived
by differentiating a first-order one). 
In addition, let the function $F$ be such that $F(\theta)> \theta$ for any $\theta$
between $0$ and $2 \pi$, whereas
$F(0)=0$ and $F(2 \pi)= 2 \pi$.  A simple cobwebbing
argument shows then that for any $\theta_0$ within the range $0 < \theta_0 < 2 \pi$,
the map generates a  discrete kink solution $..., \theta_{-1}, \theta_0, \theta_1, ...$. Therefore 
we have a one-parameter family of kinks and so Eq.\eqref{stat} represents an exceptional
discretisation.  

 This observation implies that we can find exceptional discretisations
of the sine-Gordon equation by 
considering a one-dimensional map
\be 
\theta_{n+1} - \theta_{n} = hH(\theta_{n+1},\theta_{n}), 
\label{map} 
\ee
where 
the function $H$ satisfies several requirements. First of all, it should satisfy the 
condition
\be 
H(\theta,\theta) = 2 \sin \frac{\theta}{2}
\label{conMap}
\ee
which ensures that  the map (\ref{map}) reduces to equation
\be
\theta_x= 2 \sin \frac{\theta}{2}
\label{Bogo}
\ee
in the continuum limit (where $\theta_{n+1}-\theta_n \to h \theta_x$). 
Equation \eqref{Bogo} is the Bogomolny equation  for the stationary 
continuum sine-Gordon theory: the sine-Gordon equation $\theta_{xx}= \sin \theta$
follows from Eq.\eqref{Bogo} by differentiation, 
while its kink solution \eqref{kink} is simultaneously 
a solution of Eq.\eqref{Bogo}. Therefore, the
condition \eqref{conMap} selects maps which generate discretisations of the 
sine-Gordon rather than some other equation.
Our second requirement is that 
$H(\theta_n, \theta_{n+1})$ should be bounded and positive for all pairs of 
$\theta_n$ and $\theta_{n+1}$  
with sufficiently small 
value of the difference $|\theta_{n+1}-\theta_n|$
(where $0 < \theta_n, \theta_{n+1}<2 \pi$).
Using this property of $H$, 
 assuming that $h$ is sufficiently small,
and invoking the implicit function theorem, 
we can show that Eq.\eqref{map} defines, for any $0<\theta_n< 2 \pi$,  a function $\theta_{n+1}=F(\theta_n)$, 
with $\theta_{n+1}> \theta_n$.
 Thus Eq.\eqref{map} will give rise to an {\it exceptional\/} discretisation
of the sine-Gordon equation.

This discretisation results from
 squaring both sides of (\ref{map}) and subtracting the square of its back-iterated copy,
\[
 \theta_{n} - \theta_{n-1} = hH(\theta_{n},\theta_{n-1}).
\]
This yields \cite{BOP}
\be 
\frac{\theta_{n+1} - 2\theta_{n} + \theta_{n-1}}{h^{2}} 
= \frac{H^{2}(\theta_{n+1},\theta_{n}) - H^{2}(\theta_{n},\theta_{n-1})}{\theta_{n+1} - \theta_{n-1}}.
 \label{theBus}
\ee
If $H$ is symmetric
[i.e. invariant under the permutation of
its arguments: $H(x,y)=H(y,x)$], the numerator in \eqref{theBus}
vanishes whenever the denominator equals zero and hence the 
right-hand side of \eqref{theBus} is nonsingular.
Thus the classification of exceptional discretisations reduces
to the classification of all symmetric functions $H(x,y)$ with the above
properties. The next section summarises results of this analysis.

\section{Rational-trigonometric discretisations}
\label{Mixed}

\subsection{$H^{2}(x,y) = {\cal F} (x) + {\cal F}(y)$}
\label{F_plus_F}

The simplest posibility is to let
 $ H^{2}(x,y) = {\cal F}(x) + {\cal F}(y)$. 
From $H(x,x)= 2 \sin (x/2)$ it follows that 
$ {\cal F}(x) = 2 \sin^2(x/2)$
and so
\be
 H^2(x,y) = 2 \sin^2 \frac{x}{2} +2 \sin^2 \frac{y}{2}. 
\label{F_p_F}
\ee
This function gives rise to 
one of Kevrekidis' discretizations \cite{K}:
\begin{subequations} 
\label{Kevrekidis}
\be  
\frac{\theta_{n+1} -2\theta_n + \theta_{n-1}}{h^2} 
= -\frac{\cos \theta_{n+1} - \cos \theta_{n-1}}{\theta_{n+1} - \theta_{n-1}},
 \ee
or, equivalently,
\be
\frac{\theta_{n+1} -2\theta_n + \theta_{n-1}}{h^2} 
= 2 \frac{ \sin \left[ (\theta_{n+1}-\theta_{n-1})/2 \right]}{\theta_{n+1}-\theta_{n-1}}
\sin \left( \frac{\theta_{n+1}+ \theta_{n-1}}{2} \right).
\ee
\end{subequations}

\subsection{$H^2(x,y) ={\cal F} (x+y) $}
\label{xpy}

Letting
$ H^2(x,y) ={\cal F} (x+y)$ and substituting into the continuum limit
condition, yields 
\be
H^2(x,y)=4 \sin^2 \frac{x+y}{4}.
\label{ss}
\ee
 This symmetric function
generates the 
discretisation of the form
\begin{equation}
\frac{\theta_{n+1} -2\theta_n + \theta_{n-1}}{h^{2}} 
= 
4 \frac{ \sin\left[ (\theta_{n+1}-\theta_{n-1})/4 \right]}{\theta_{n+1}-\theta_{n-1}}
\sin \left( \frac{\theta_{n+1}+ 2 \theta_n+ \theta_{n-1}}{4} \right).
\label{Z1}
\end{equation}

\subsection{ $H^2(x,y) = {\cal F}(x) {\cal F}(y)$ \label{first}}

Another simple possibility is to assume that 
$ H^2(x,y)={\cal F}(x)  {\cal F}(y)$. From the continuum limit we obtain $ {\cal F}(x)= 2 \sin (x/2)$ and so
\be 
H^2(x,y) =4 \sin \frac{x}{2} \sin \frac{y}{2}.
\label{s}
\ee
 By substituting into equation (\ref{theBus}) we find 
the following discretisation:
\be \frac{\theta_{n+1} -2\theta_n + \theta_{n-1}}{h^{2}} 
= 8  \frac{\sin[(\theta_{n+1}-\theta_{n-1})/4]}
{\theta_{n+1}- \theta_{n-1}} \sin  \left( \frac{\theta_n}{2} \right)
  \cos \left(
\frac{\theta_{n+1}+\theta_{n-1}}{4}
\right). 
\label{Z2}
\ee

\subsection{$H^2(x,y) =[ {\cal F}(x)+ {\cal F} (y)]^2$}
\label{FxFy}

Considering the symmetric function of the form
$ H^2(x,y)=[  {\cal F}(x)+ {\cal F} (y)]^2$, 
we obtain from the continuum limit: 
\be
H^2(x,y)= \left( \sin \frac{x}{2} + \sin \frac{y}{2} \right)^2.
\label{H_FxFy}
\ee
This gives rise to the following discretisation:
\begin{eqnarray}
\frac{\theta_{n+1} -2\theta_n + \theta_{n-1}}{h^{2}} 
= 2 
\frac{ \sin\left[ (\theta_{n+1}-\theta_{n-1})/4 \right]}{\theta_{n+1}-\theta_{n-1}}
\cos \left( \frac{\theta_{n+1}+ \theta_{n-1}}{4} \right) \nonumber \\
\times \left( \sin \frac{\theta_{n+1}}{2} +
2 \sin \frac{\theta_n}{2} 
+ \sin \frac{\theta_{n-1}}{2} 
\right).
\label{Z3}
\end{eqnarray}

\subsection{\label{third}$H^{2}(x,y) =  {\cal F}(x) {\cal F}(y) +  {\cal G}(x) {\cal G}(y)$}
\label{FG}

A simple symmetric generalisation involving two functions of a single argument,
say $ {\cal F}(x)$ and $ {\cal G}(x)$, is 
 $ H^{2}(x,y) = {\cal F}(x){\cal F}(y) + {\cal G}(x){\cal G}(y)$. 
Setting $x=y$ yields
\[
{\cal F}^2(x) + {\cal G}^2(x) = 4\sin^{2}\frac{x}{2}.  
\]
One possibility here is to assume that the functions ${\cal F}$ and ${\cal G}$ have the form
\[
 {\cal F}(x) = 2\eta(x)\sin\frac{x}{2}, \quad {\cal G}(x) = 2\xi(x)\sin\frac{x}{2}, 
\]
where  $\eta$ and $\xi$ satisfy $\eta^2(x) + \xi^2(x) = 1$.
The simplest trigonometric choice for $\eta$ and $\xi$ is
\[
 \eta(x) = \sin(ax), \quad
\xi(x) = \cos(ax), 
\]
where $a$ is a  parameter. Taking, for instance, $a = \frac{1}{2}$,
 gives us the following expression for $H^{2}(x,y)$:
\be 
H^{2}(\theta_{n+1}, \theta_n) = 4
\sin \frac{\theta_{n+1}}{2}
\sin \frac{\theta_n}{2}
\cos \frac{\theta_{n+1}-\theta_n}{2}. 
\label{HFG}
\ee
The function \eqref{HFG} is obviously positive for
 $|\theta_{n+1}-\theta_n| <\pi$ and
therefore the resulting discretization,
\be 
\frac{\theta_{n+1} -2\theta_n + \theta_{n-1}}{h^{2}}
= 4 \frac {\sin \left[ (\theta_{n+1}- \theta_{n-1})/2 \right]} {\theta_{n+1} - \theta_{n-1}}
 \sin \left( \frac{\theta_n}{2} \right)
\cos \left( \frac{\theta_{n+1}-\theta_n+ \theta_{n-1}}{2} \right),
\label{Z4}
\ee
is exceptional for sufficiently small $h$.

Another simple symmetric combination of two functions of a single argument, is
$H^2(x,y)={\cal F}(x){\cal G}(y)+{\cal F}(y){\cal G}(x)$; however, this $H^2$ gives rise to the 
discretisation that we have already identified, Eq.\eqref{Kevrekidis}. 

\subsection{More complex symmetric functions}
It is not difficult to construct more examples of symmetric functions 
$H(x,y)$, with increasing complexity. One possibility is to
take
\[
H^2(x,y)= \sum_{n=1}^N {\cal F}_n(x) {\cal F}_n(y),
\]
where ${\cal F}_n(x)$ ($n=1,2,...,N$) are appropriate trigonometric functions. Another
symmetric combination is
\[
H^2(x,y)=  
\prod_{n=1}^N {\cal F}_n(x_{|n|}) +
\prod_{n=1}^N {\cal F}_n(x_{|n+1|}),
\]
where $|n| \equiv n \ {\rm mod} \, 2$, and $x_1 =x$, $x_2=y$.

\section{Purely trigonometric discretisations}
\label{Pure}

Our original one-dimensional map (\ref{map})
can be modified  to produce new, periodic, discretizations. 
Instead of \eqref{map}, we consider the map
\be 
\label{kMap} 
\ell \sin \frac{\theta_{n+1} - \theta_n}{\ell} = hH(\theta_{n+1}, \theta_n),
\ee
where $H(\theta_{n+1}, \theta_n)$ is a trigonometric function of its arguments
and $\ell$ is a positive integer.
Subtracting from the square of \eqref{kMap} 
the square of its back-iterated copy yields the
discrete model
\be 
\label{theBus2} 
\frac{\ell^2}{h^2} \sin \frac{\theta_{n+1} - 2\theta_n + \theta_{n-1}}{\ell} 
= \frac{ H^2(\theta_{n+1},\theta_n) - H^2(\theta_n,\theta_{n-1}) }
 {\sin [(\theta_{n+1} - \theta_{n-1})/\ell]}.
\ee
As in Eq.\eqref{map}, we assume that 
$H(\theta_{n+1}, \theta_n)$ 
is positive, symmetric and has the continuum limit $H(\theta,\theta)=2 \sin(\theta/2)$.
For sufficiently small $h$ and  $|\theta_{n+1}-\theta_n|$, Eq.\eqref{kMap} defines an
implicit function $\theta_{n+1}=F(\theta_n)$,
with $\theta_{n+1}> \theta_n$. Consequently, the discretisation \eqref{theBus2}
is exceptional.

The
 discretisations \eqref{theBus2} 
are different from those in \eqref{theBus}  in that 
every term in \eqref{theBus2} is periodic in
each of its three arguments, $\theta_{n-1}$, $\theta_n$ and $\theta_{n+1}$. 
The models of the form \eqref{theBus2} 
find their applications in the description of coupled chains of elements 
where each element is characterised by a periodic variable (an angle)
and the coupling of elements does not violate this periodicity.
One example is given by the Speight-Ward  discretisation \cite{SW}
\begin{equation}
\frac{4}{h^2}
\sin \left(
\frac{\theta_{n+1} -2\theta_n + \theta_{n-1}}{4} \right)
= 
\sin \left( \frac{\theta_{n+1}+ 2 \theta_n+ \theta_{n-1}}{4} \right),
\label{SpeWar}
\end{equation}
which
has recently been shown to describe 
 chains of electric dipoles constrained to rotate in the plane containing the chain \cite{Speight_Zolo}.
Another example is the one-dimensional chiral $XY$ model
\begin{subequations}
\label{XY}
\be
\sin(\chi_{n+1}-\chi_n-\gamma)  - \sin(\chi_n - \chi_{n-1}-\gamma)= K \sin(p \chi_n) \quad (p=1,2,...),
\ee
or, equivalently,
\be
\frac{\ell}{h^2} \sin \left( \frac{\theta_{n+1}- 2\theta_n + \theta_{n-1}}{\ell} \right)
\cos \left( \frac{\theta_{n+1}- \theta_{n-1}}{\ell} -\gamma \right) = \sin \theta_n,
\ee
\end{subequations}
where $\theta_n=p \chi_n$, $\ell=2p$ and $h^2=pK$.
The chiral $XY$ model \eqref{XY} describes arrays of spins with the nearest-neighbour interactions 
in an external magnetic field \cite{chiral_XY,Grif}.
It is used
to model helimagnetic materials, discotic and ferroelectric 
smectic liquid crystals, crystalline polymers,  
thin magnetic films and Josephson junction arrays.

The classification of discretisations of the form \eqref{theBus2} reduces
to the classification of all possible symmetric functions $H(x,y)$ --- 
the task completed in section \ref{Mixed} above. 
Each function $H^2(x,y)$ identified in section \ref{Mixed}
gives rise to a number of purely periodic discretisation of the form \eqref{theBus2},
with various $\ell$;
that is, each rational-trigonometric exceptional model \eqref{theBus}
has a set of purely trigonometric counterparts \eqref{theBus2}. 
We will restrict ourselves to the simplest representative(s) of these sets
by choosing  appropriate value(s) of 
$\ell$. The resulting models can be summarised as follows.

Picking
the symmetric function \eqref{F_p_F} of the section \ref{F_plus_F}
and letting $\ell=2$,
gives rise to a very simple exceptional discretisation 
\be 
\frac{2}{h^2} \sin \left(
\frac{\theta_{n+1} -2\theta_n + \theta_{n-1}}{2} \right)
= \sin \left( \frac{\theta_{n+1}+ \theta_{n-1}}{2} \right).
\label{A}
\ee
If, instead, we took $\ell=4$, we would obtain a slightly
more complicated model:
\be 
\frac{4}{h^2} \sin \left(
\frac{\theta_{n+1} -2\theta_n + \theta_{n-1}}{4} \right)
=  \sin \left( \frac{\theta_{n+1}+ \theta_{n-1}}{2} \right)
\cos \left( \frac{\theta_{n+1}-\theta_{n-1}}{4}\right).
\label{AAA}
\ee
Finally, if we ``extend" the symmetric function \eqref{F_p_F} by adding a term that 
vanishes in the continuum limit,
\[
H^2(x,y)= 2 \sin^2 \frac{x}{2} + 2 \sin^2 \frac{y}{2} + 2 \sin^2 \frac{x-y}{2},
\]
then, keeping $\ell=4$, we will arrive at a
(still reasonably simple) exceptional model
\be 
\frac{2}{h^2} \sin \left(
\frac{\theta_{n+1} -2\theta_n + \theta_{n-1}}{4} \right)
=  \sin \left( \frac{\theta_{n+1}- \theta_n + \theta_{n-1}}{2} \right)
\cos \left( \frac{\theta_{n+1}-\theta_{n-1}}{4}\right) \cos \frac{\theta}{2}.
\label{AA}
\ee

Next, choosing the symmetric function \eqref{ss} of the section
\ref{xpy} and letting $\ell=4$ yields  Speight and Ward's model, Eq.\eqref{SpeWar}.
On the other hand, taking the symmetric function \eqref{s} of  section \ref{first}
and letting $\ell=4$, gives the discretisation 
\be 
\frac{4}{h^2} \sin \left(
\frac{\theta_{n+1} -2\theta_n + \theta_{n-1}}{4} \right)
= 2   \sin  \left( \frac{\theta_n}{2} \right)
  \cos \left(
\frac{\theta_{n+1}+\theta_{n-1}}{4}
\right).
\label{plu_min}
\ee
Eq.\eqref{plu_min} reduces to Eq.\eqref{SpeWar}
with the lattice spacing constant ${\tilde h}=h(1+h^2/4)^{-1/2}$, 
if we use an identity 
\begin{equation}
2 \sin  \left( \frac{\theta_n}{2} \right) \cos  \left(\frac{\theta_{n+1} + \theta_{n-1}}{4} \right)=
\sin \left( \frac{\theta_{n+1} + 2 \theta_n + \theta_{n-1}}{4} \right)
-  \sin \left( \frac{\theta_{n+1}- 2 \theta_n + \theta_{n-1}}{4} \right).
\label{id} 
\end{equation}  

Picking the symmetric function of the form \eqref{H_FxFy} from
section \ref{FxFy}, and letting $\ell=4$, gives an exceptional discretisation
\be
\frac{4}{h^2}
\sin \left(
\frac{\theta_{n+1} -2\theta_n + \theta_{n-1}}{4} \right)
= \frac12 
\cos \left( \frac{\theta_{n+1}+ \theta_{n-1}}{4} \right) 
 \left( \sin \frac{\theta_{n+1}}{2} +
2 \sin \frac{\theta_n}{2} 
+ \sin \frac{\theta_{n-1}}{2} 
\right).
\label{dd30}
\ee

Finally, choosing the symmetric function in the form
\eqref{HFG} from section \ref{FG}
and letting $\ell=2$ produces an exceptional model
\be 
\frac{2}{h^2} \sin \left(
\frac{\theta_{n+1} -2\theta_n + \theta_{n-1}}{2} \right)
= 2
  \sin \left( \frac{\theta_n}{2} \right)
\cos \left( \frac{\theta_{n+1}-\theta_n+ \theta_{n-1}}{2} \right).
\ee
Writing the right-hand side as a sum of sines, we reproduce equation \eqref{A}
with $h$ replaced with ${\tilde h}=h(1+h^2/2)^{-1/2}$.

The models \eqref{A}, \eqref{AAA}, \eqref{AA} and \eqref{dd30} constitute
our list of new exceptional periodic discretisations of the sine-Gordon equation.
This list  
can be generalised and extended in a variety of ways.
For example, we can replace the sine function in \eqref{kMap} with 
$ \tan [(\theta_{n+1}-\theta_n)/\ell]$
or, more generally, with $ \sin[(\theta_{n+1}-\theta_n)/\ell] \cos^p [m(\theta_{n+1}-\theta_n)]$
with arbitrary $m$ and $p$. 
Also,  we can add a
sum $\sum A_n \sin^2[B_n(x-y)]$ with arbitrary $A_n$ and $B_n$
to any of the symmetric functions $H^2(x,y)$. Since we are mainly interested in simple discretisations, we are not
pursuing these possibilities in our present work.

\section{Discrete sine-Gordon equation with exact kink solutions}
\label{Exact}

In this section we  consider
one more exceptional discretisation of the sine-Gordon
equation. In addition to 
admitting an  arbitrary centring relative to the lattice, the kinks 
in this model are available in exact explicit form.

We start with what may seem to be an unrelated map,
\[
\phi_{n+1} -\phi_n=h(1- \phi_n \phi_{n+1}).
\]
Writing the square of this map as
\[
\frac{(\phi_{n+1} -\phi_n)^2}{1- \phi_{n+1} \phi_n}=h^2(1- \phi_n \phi_{n+1}),
\]
and subtracting its back-iterated copy gives
\begin{subequations} \label{d300}
\begin{equation}
 \phi_{n+1}-2 \phi_n +\phi_{n-1} +  \phi_n (\phi_n^2-\phi_{n+1} \phi_{n-1})= - h^2 \phi_n(1-\phi_{n+1}\phi_n)(1-\phi_n \phi_{n-1}).
\label{d3}
\end{equation}
Equation \eqref{d3} with 
\begin{equation}
\phi_n= \cos \frac{\theta_n}{2},  
\label{subs}
\end{equation}
\end{subequations}
that is, equation
\begin{eqnarray}
\cos \frac{\theta_{n+1}}{2}- 2 \cos \frac{\theta_n}{2} + \cos \frac{\theta_{n-1}}{2}
+ \cos \frac{\theta_n}{2} 
\left( \cos^2 \frac{\theta_n}{2}- \cos \frac{\theta_{n+1}}{2} \cos \frac{\theta_{n-1}}{2} \right)
 \nonumber \\
=-h^2 \cos \frac{\theta_n}{2}
\left( 1 - \cos \frac{\theta_n}{2} \cos \frac{\theta_{n+1}}{2} \right)
\left( 1 - \cos \frac{\theta_n}{2} \cos \frac{\theta_{n-1}}{2} \right),
\label{complex_sine_Gordon}
\end{eqnarray}
provides an exceptional discretisation of the sine-Gordon
equation.
Indeed, the continuum limit of  equation \eqref{d3} is 
\begin{equation}
\phi_{xx} (1-\phi^2) + \phi \phi_x^2 = -\phi(1- \phi^2)^2,
\label{contin_1}
\end{equation}
which is nothing but the stationary sine-Gordon equation $\theta_{xx}= \sin \theta$
 written in terms of
 $\phi= \cos (\theta/2)$.

Equation \eqref{d3} has an exact kink solution 
\[
\phi_n= {\rm tanh} (kn-x^{(0)}), \quad \tanh k=h,
\]
where $x^{(0)}$ is a translation parameter which can be chosen arbitrarily.
Applying the transformation \eqref{subs} to $\phi_n$ produces an explicit kink solution
of the discrete sine-Gordon equation \eqref{complex_sine_Gordon}:
\begin{equation}
\theta_n=  4 \arctan [\exp (kn-x^{(0)})], \quad \tanh k=h.
\label{d60}
\end{equation}

We are not aware of any physical systems represented by Eq.\eqref{complex_sine_Gordon}.
This discrete model may find its uses, however, in numerical simulations of the continuum sine-Gordon equation.
Like other exceptional discretisations of the sine-Gordon equation, 
this model preserves an ``effective translation  invariance" of the continuum equation.
The fact that  the stationary discrete kinks of the model \eqref{complex_sine_Gordon} are available 
in exact explicit form  is an additional computational
advantage.

\section{Travelling kinks}
\label{Travelling}

In this section we show how the method of one-dimensional maps can be 
used to construct moving kinks. 

The discretisation breaks the Lorentz invariance of the continuum model \eqref{sG}
in the same way as it breaks its translation symmetry; hence the mobility of the kink becomes
a nontrivial property in the discrete case. As the kink moves in the Peierls-Nabarro potential, 
it excites resonant radiation and decelerates as a result of that \cite{Currie,Willis_1986,Ishimori_and_Peyrard,K0,KW}.
Surprisingly, some discrete models exhibit isolated values of the kink
velocity for which the kink can {\it slide\/}, 
i.e. travel without experiencing radiative friction
\cite{sliding,FZK,OPB,Zakrzewski}. 

 A pertinent question here is whether there are 
exceptional discretisations where the kink can slide with an {\it arbitrary\/}
velocity. A 
  discrete nonlinear Schr\"odinger equation with this property is well
known; 
it is the Ablowitz-Ladik model whose solitons are radiationless
irrespective of their velocities.
On the other hand, no discrete Klein-Gordon equations whose  kink velocities would {\it all\/}
 be sliding velocities have been found so far --- neither in the Frenkel-Kontorova class of models nor
among the discrete $\phi^4$-theories.

In this section we construct such a discrete sine-Gordon equation. 
Its kink solutions are given by explicit expressions,
and, as  will become obvious from these explicit formulas, 
all its kinks travel without emitting radiation.

We start with a nonstationary  equation
\begin{equation}
\phi_{xx}-\phi_{tt} -2 \phi \frac{\phi_x^2-\phi_t^2}{1+\phi^2}  = \phi
\frac{1- \phi^2}{1+\phi^2},
\label{d27}
\end{equation}
which transforms into the  sine-Gordon equation $\theta_{xx}-\theta_{tt}= \sin \theta$
by the substitution $
\phi= \tan (\theta/4)$. 
Our first observation is that if 
 $\phi(x,t)$ is a simultaneous solution of two first-order equations
\begin{equation}
\phi_x= \frac{1}{\sqrt{1-v^2}} \phi  
\label{d28}
\end{equation}
and
\begin{equation}
\phi_t= -\frac{v}{\sqrt{1-v^2}} \phi,
\label{d29}
\end{equation}
then it also satisfies Eq.\eqref{d27}.
In  \eqref{d28} and \eqref{d29}, $v$ is a parameter; $-1<v<1$.
Note that Eq.\eqref{d27} does not contain $v$ explicitly; hence 
finding a solution of equations \eqref{d28} and \eqref{d29} for {\it all\/}
$v$ amounts to finding a one-parameter family of solutions to
\eqref{d27}.

Next, we discretise Eq.\eqref{d28} according to 
\begin{equation}
\frac{1}{h^2} (\phi_{n+1} - \phi_n)^2= \frac{1}{1-v^2} \phi_n \phi_{n+1}
\label{d30}
\end{equation}
and divide both sides by the same expression to get
\begin{equation}
\frac{1}{h^2} \frac{(\phi_{n+1} - \phi_n)^2}
{(1+ \phi_{n+1}^2) (1+ \phi_{n}^2)}
= \frac{1}{1-v^2} \frac{\phi_n \phi_{n+1}}{(1+ \phi_{n+1}^2) (1+
\phi_{n}^2)}.
\label{d30p}
\end{equation}
We also consider a discrete version of Eq.\eqref{d29}:
\begin{equation}
{\dot \phi_n}= - \frac{v}{\sqrt{1-v^2}} \phi_n.
\label{starr}
\end{equation}
Subtracting Eq.\eqref{d30p}  from its back-iterated copy
and replacing 
 $\phi_n$ with $\tan (\theta_n/4)$,
 gives 
\begin{equation}
  \frac{1}{h^2} \sin \left( \frac{\theta_{n+1}-2 \theta_n + \theta_{n-1}}{4} \right) =
\frac12 \, \frac{1}{1-v^2} \sin \left( \frac{\theta_n}{2} \right)
 \cos  \left( \frac{\theta_{n+1} + \theta_{n-1}}{4} \right).
\label{d31}
\end{equation}
On the other hand, Eq.\eqref{starr} yields
\begin{equation}
{\ddot \phi_n} 
-2 \phi_n  \frac{{\dot \phi_n}^2}{1+ \phi_n^2}
=\frac{v^2}{1-v^2} \phi_n \frac{1- \phi_{n-1} \phi_{n+1}}{ 1 + \phi_{n-1} \phi_{n+1}},
\label{d32}
\end{equation}
where we have used the relation $\phi_{n+1} \phi_{n-1} = \phi_n^2$ which is straightforward from
\eqref{d30}. Letting $\phi_n= \tan (\theta_n/4)$  in \eqref{d32}, we get
\begin{equation}
\frac{\ddot \theta_n}{4}
 \cos \left( \frac{\theta_{n+1}-\theta_{n-1}}{4} \right)
 =
\frac12 \,
\frac{v^2}{1-v^2} 
\sin \left( \frac{\theta_n}{2} \right)
\cos \left( \frac{\theta_{n+1}+ \theta_{n-1}}{4} \right).
\label{d33}
\end{equation}
Finally, subtracting \eqref{d31}  from \eqref{d33} yields
 a discrete sine-Gordon equation
\begin{equation}
\cos  \left( \frac{ \theta_{n+1}- \theta_{n-1}}{4} \right) \frac{\ddot \theta_n}{4} =
\frac{1}{h^2}  \sin \left( \frac{\theta_{n+1}-2 \theta_n + \theta_{n-1}}{4} \right)-
\frac{1}{2} \sin  \left( \frac{\theta_n}{2} \right)
\cos \left( \frac{\theta_{n+1} + \theta_{n-1}}{4} \right).
\label{d34}
\end{equation}
For any $h$, this equation has an explicit moving kink solution which
is a compatible solution of the first-order difference equation
\eqref{d30} and the first-order differential equation \eqref{starr}:
\begin{equation}
\theta_n = 4 \arctan \left[ \exp \left( kn -\frac{vt}{\sqrt{1-v^2}} \right) \right],
\label{d35}
\end{equation}
where  $k$ is defined by 
\begin{equation}
 2 \sinh \left( \frac{k}{2}  \right) = \frac{h}{\sqrt{1-v^2}}
\label{d36}
\end{equation} 
and $v$ can take any value between $-1$ and $1$.
As $h \to 0$, the solution \eqref{d35}-\eqref{d36} tends to the 
travelling kink solution
of the continuum sine-Gordon equation, 
\[
\theta_n \to 4 \arctan \left[ \exp \left( \frac{x_n-vt}{\sqrt{1-v^2}} \right) \right],
\quad x_n=hn.
\]

Using an identity \eqref{id}, equation \eqref{d34} can be cast in the form
\begin{equation}
\cos  \left( \frac{ \theta_{n+1}- \theta_{n-1}}{4} \right) {\ddot \theta_n} =
 \frac{4}{{\tilde h}^2}  \sin \left( \frac{\theta_{n+1}-2 \theta_n + \theta_{n-1}}{4} \right)-
\sin  \left( \frac{\theta_{n+1} + 2 \theta_n + \theta_{n-1}}{4}
\right),
\label{d37}
\end{equation}
where ${\tilde h}=h(1+h^2/4)^{-1/2}$.
Solution to \eqref{d37} is given by the same Eq.\eqref{d35} where
$k$ should now be defined by 
\begin{equation}
   \sinh \left( \frac{k}{2}  \right) = \frac{1}{\sqrt{1-v^2}} \frac{\tilde h}{\sqrt{4-{\tilde h}^2}}.
\label{d38}
\end{equation}
Solution \eqref{d35},\eqref{d38} exists for any $|v|<1$ and  $0<{\tilde h}<2$.

We close this section by noting
that the stationary limit of Eq.\eqref{d37} coincides with the
stationary part of the Speight-Ward model \cite{SW},
\begin{equation}
\cos^{-1} \left( \frac{\theta_{n+1}-\theta_{n-1}}{4} \right)
{\ddot \theta_n}  =
 \frac{4}{{\tilde h}^2}  \sin \left( \frac{\theta_{n+1}-2 \theta_n + \theta_{n-1}}{4} \right)-
\sin  \left( \frac{\theta_{n+1} + 2 \theta_n + \theta_{n-1}}{4}
\right).
\label{SW_nonstationary}
\end{equation}
(Here $\cos^{-1} \alpha$ should be understood as $1/ \cos \alpha$ and
not as $\arccos \, \alpha$.) 
The simulations of Speight and Ward
\cite{SW} have demonstrated that the motion of the kink in their
Eq.\eqref{SW_nonstationary} is accompanied by a much weaker radiation than 
the kink propagation 
in a typical nonexceptional model. Now that we have another time-dependent version of the same stationary model, 
in which the radiation is completely suppressed for all velocities, 
the low level of radiation from the moving Speight-Ward kink 
can be explained simply by the proximity of their equation \eqref{SW_nonstationary}
to our model \eqref{d37}.

\section{Concluding remarks}
\label{Conclusions}

Results of this work can be summarised as follows.

\begin{itemize}
\item
Using the method of one-dimensional maps, we have derived 
several exceptional discretisations 
of the sine-Gordon equation involving ratios of trigonometric
to linear functions: equations \eqref{Z1}, \eqref{Z2}, \eqref{Z3} and \eqref{Z4}.
All these exceptional models are new. We have also recovered
the exceptional system of Kevrekidis, Eq.\eqref{Kevrekidis}, which was originally 
obtained within a different approach \cite{K}.

\item
We have identified several new purely-trigonometric exceptional 
discretisations, in particular
equations \eqref{A}, \eqref{AAA} and \eqref{dd30}:
\[
{\ddot \theta_n}=
\frac{2}{h^2} \sin \left(
\frac{\theta_{n+1} -2\theta_n + \theta_{n-1}}{2} \right)
- \sin \left( \frac{\theta_{n+1}+ \theta_{n-1}}{2} \right);
\]
\[ 
{\ddot \theta_n}=
\frac{4}{h^2} \sin \left(
\frac{\theta_{n+1} -2\theta_n + \theta_{n-1}}{4} \right)
-  \sin \left( \frac{\theta_{n+1}+ \theta_{n-1}}{2} \right)
\cos \left( \frac{\theta_{n+1}-\theta_{n-1}}{4}\right);
\]
\[
{\ddot \theta_n}=
\frac{4}{h^2}
\sin \left(
\frac{\theta_{n+1} -2\theta_n + \theta_{n-1}}{4} \right)
- \frac12 
\cos \left( \frac{\theta_{n+1}+ \theta_{n-1}}{4} \right) 
 \left( \sin \frac{\theta_{n+1}}{2} +
2 \sin \frac{\theta_n}{2} 
+ \sin \frac{\theta_{n-1}}{2} 
\right).
\]

\item
We have derived a new discretisation with exact explicit kink solutions, 
Eq.\eqref{complex_sine_Gordon}.

\item
We have constructed a discrete sine-Gordon model which 
supports kinks travelling with arbitrary velocities:
\be
{\ddot \theta_n} \cos  \left( \frac{ \theta_{n+1}- \theta_{n-1}}{4} \right) =
 \frac{4}{h^2}  \sin \left( \frac{\theta_{n+1}-2 \theta_n + \theta_{n-1}}{4} \right)-
\sin  \left( \frac{\theta_{n+1} + 2 \theta_n + \theta_{n-1}}{4}
\right).
\label{moving}
\ee
\end{itemize}

The latter result deserves an additional comment.
By analogy
with the derivation of the model \eqref{moving}, it
 is not difficult to construct discrete $\phi^4$ theories supporting 
sliding kinks with arbitrary velocities. One such model has
the form
\begin{equation}
{\ddot \phi_n} \frac{1-\phi_{n+1} \phi_{n-1}}{1-\phi_n^2}=
 \frac{\phi_{n+1}-2 \phi_n + \phi_{n-1}}{h^2} +
\frac{\phi_n}{2} ( 1-\phi_{n+1} \phi_{n-1}).
\label{moving_4}
\end{equation}
(This is a time-dependent generalisation of the exceptional stationary $\phi^4$
model derived in \cite{BOP}.)
The moving kink solution to Eq.\eqref{moving_4} has the form
\[
\phi_n= {\rm tanh} \, \left( kn- \frac{vt}{2\sqrt{1-v^2}} \right),
\quad 
\frac{4\tanh^2 k}{1+  \tanh^2 k}=\frac{h^2}{1-v^2}.
\]
Another time-dependent discretisation of the $\phi^4$ theory
with sliding kinks is
\begin{equation}
{\ddot \phi_n} \frac{1-\phi_{n+1} \phi_{n-1}}{1-\phi_n^2}=
 \frac{\phi_{n+1}-2 \phi_n + \phi_{n-1}}{h^2} +
\frac{\phi_n}{2} \left[ 1-\frac{\phi_n}{2} (\phi_{n+1} +\phi_{n-1})\right].
\label{moving_5}
\end{equation}
(This is a time-dependent generalisation of the  exceptional 
$\phi^4$ model identified by Bender and Tovbis \cite{BT} and Kevrekidis \cite{K}.) 
The sliding kink solution has the form
\[
\phi_n= \tanh \, \left( kn- \frac{1}{2 \sqrt{1+ \tanh^2 k}} \frac{vt}{\sqrt{1-v^2}} \right),
\quad 
4\tanh^2 k=\frac{h^2}{1-v^2}.
\]

The analogy between equation \eqref{moving} and the Ablowitz-Ladik 
model is also worth
commenting upon. The Ablowitz-Ladik  model is the only discrete nonlinear
Schr\"odinger equation whose solitons can slide with any chosen
velocity. The absence of the accompanying radiation is usually explained by the
integrability
of this equation. A similar
behaviour of kinks of Eq.\eqref{moving} makes one wonder whether the latter equation
could also be integrable. We have tested the integrability of
Eq.\eqref{moving} numerically,
by simulating a collision of a kink and an antikink.
The scattering was found to be inelastic: the velocities of the kink
and antikink changed as a result of the collision, 
and significant amount of radiation was detected. Consequently, we
conclude that equation \eqref{moving} is not integrable. This 
example demonstrates that, contrary to common belief, the
integrability
is not a prerequisite for the existence of a discrete soliton sliding 
at an arbitrarily chosen velocity.

 Finally, it is interesting to compare travelling kink solutions of
our model \eqref{moving} with travelling kinks of another modification of the 
Speight-Ward model proposed by Zakrzewski \cite{Zakrzewski}. 
Zakrzewski's model is different from the Speight-Ward equation \eqref{SW_nonstationary}
in the presence of a factor $(1+ \alpha {\dot \theta_n}^2)^{-1}$ 
in front of the left-hand side of \eqref{SW_nonstationary}, with $\alpha= const$.
 The model has an
exact solution in the form of a sliding kink; however, similarly
to radiationless moving kinks in other systems \cite{sliding,FZK,OPB}, this kink 
can only slide with one particular velocity which is determined by
the parameters $h$ and $\alpha$. Unlike this codimension-1 
solution, our sliding kinks \eqref{d35}, \eqref{d38} have codimension 0 in the
sense that they can move with an arbitrary velocity independent of $h$.

\acknowledgments
 IB was supported by the NRF of South Africa under grant 2053723.
TvH was supported by the National Institute of Theoretical Physics.

\end{document}